\def\etbrcl{$\kappa$-(BE\-DT\--TTF)$_2$\-Cu\-[N\-(CN)$_{2}$]\-Br$_{x}$Cl$_{1-x}$}
\def\cm{cm$^{-1}$}
\documentclass[twocolumn,prl,aps]{revtex4}
\usepackage{graphics}
\usepackage{dcolumn}
\usepackage{bm}
\usepackage{amsmath}
\usepackage{amssymb}
\usepackage{epsfig}
\usepackage{pictex}
\begin{document}

\title{Quasiparticles at the verge of localization near
the Mott \\ metal-insulator transition in a two-dimensional
material}
\author{J. Merino$^{1}$, M. Dumm$^{2}$, N. Drichko$^{2,3}$, M. Dressel$^{2}$,
and Ross H. McKenzie$^{4}$} \affiliation{$^1$ Departamento de
F\'isica Te\'orica de la Materia
Condensada, Universidad Aut\'onoma de Madrid, Madrid 28049, Spain\\
$^2$ 1.~Physikalisches Institut, Universit\"at Stuttgart,
Pfaffenwaldring 57, 70550 Stuttgart, Germany\\
$^3$ Ioffe Physico-Technical Institute, Russian Academy of
Science,
194021 St.Petersburg, Russia\\
$^4$ Physics Department, University of Queensland,
 Brisbane 4072, Australia}
\date{\today}
\bigskip
\begin{abstract}
The dynamics of charge carriers close to the Mott transition is
explored theoretically and experimentally in the quasi
two-dimensional organic charge-transfer salt,
$\kappa$-(BEDT-TTF)$_2$\-Cu[N(CN)$_2$]\-Br$_x$Cl$_{1-x}$, with
varying Br content. The frequency dependence of the conductivity
deviates significantly from simple Drude model behavior: there is
a strong redistribution of spectral weight as the Mott transition
is approached and with temperature. The effective mass of the
quasiparticles increases considerably when coming close to the
insulating phase. A dynamical mean-field-theory treatment of the
relevant Hubbard model gives a good quantitative description of
the experimental data.
\end{abstract}
\pacs{
71.30.+h, 
74.25.Gz, 
71.10.Hf, 
74.70.Kn   
}

\maketitle
\bigskip

Understanding strongly correlated electron materials such as the
high-$T_c$ cuprate superconductors, heavy fermions, transition
metal oxides, and organic superconductors is a significant
theoretical challenge.
 Optical conductivity measurements
are a powerful probe of the competition between the itineracy and
localization of charge carriers
\cite{Marel05,basov,degiorgi,DresselGruner02}.
 Unlike in elemental metals, a
large redistribution of spectral weight (SW) occurs with variation
in the Coulomb interaction and/or
temperature\cite{Rozenberg,basov,degiorgi,Jacobsen,Vescoli98}.
 The simple Drude model \cite{DresselGruner02} fails to describe spectra in these
systems, despite its success in conventional metals.
 In this Letter, we present
a combined theoretical and experimental analysis of the
redistribution of SW that occurs as the Mott metal-insulator
transition is approached in a two-dimensional material.

The $\kappa$-family of layered organic superconductors
 based on the
bis-(ethylenedithio)tetrathiafulvalene (BEDT-TTF) molecules
\cite{Ishiguro01}
 have conduction bands which are effectively half filled (due to
dimerization of neighboring BEDT-TTF molecules). As the on-site
Coulomb repulsion $U$ is comparable to the bandwidth $W$, these
materials
 are ideal systems to explore the bandwidth controlled
Mott metal-insulator transition (MIT) in two dimensions.
 Measurements of the DC conductivity have been made as the materials
tuned  through the MIT via chemical substitution, pressure,
magnetic field, and/or temperature \cite{Limelette00,Kanoda05}.
Much less is known about the charge-carriers dynamics in these
systems close to the MIT
\cite{Jacobsen,Eldridge91,Sasaki04,Nam07}.
 The optical conductivity calculated from
a dynamical mean-field theory (DMFT)\cite{Kotliar96} treatment of
the relevant Hubbard model
 is compared with optical spectra of single crystals of
$\kappa$-(BEDT-TTF)$_2$Cu[N(CN)$_2$]Br$_x$Cl$_{1-x}$. Partial
substitution of Br  by Cl in the anion layer increases the
effective Coulomb repulsion $U/W$, and drives the system
 across the metal-insulator phase boundary, which
is at $x \simeq 0.7$ \cite{Faltermeier07}.

The simplest strongly correlated electron model for the conducting
layers of the $\kappa$-(BEDT-TTF)$_2X$ family is a Hubbard model
on a frustrated square lattice
 at half filling \cite{powell}:
\begin{equation}
H =-t_2 \sum_{\langle ij \rangle,\sigma} (c^\dagger_{i \sigma}
c_{j \sigma} + c^\dagger_{j \sigma} c_{i \sigma})
-t_1\sum_{\langle\langle ij \rangle\rangle,\sigma} (c^\dagger_{i
\sigma} c_{j \sigma} + c^\dagger_{j \sigma} c_{i \sigma})
\nonumber
\end{equation}\vspace*{-5mm}
\begin{equation}
+ U \sum_{i} n_{i\uparrow} n_{i\downarrow} -\mu \sum_{i \sigma}
c^{\dagger}_{i\sigma} c_{i \sigma} \quad , \label{model}
\end{equation}
where $\sigma$ is the spin index, $t_1$ describes hopping along
one diagonal of the square plaquettes, $t_2$ describes hopping
around the plaquettes, and $c^\dagger_{i \sigma}$ and $c_{i
\sigma}$ are hole creation and destruction operators on the
antibonding orbitals of BEDT-TTF dimers.
 The non-interacting model ($U=0$) has a dispersion
relation:
\begin{equation}
\epsilon({\bf k}) = -2t_2[\cos(k_x d)+\cos(k_y
d)]-2t_1\cos((k_x+k_y)d)\label{tight}
\end{equation}
 with $d$ the nearest-neighbor dimer distance.
 We take the nearest-neighbor hopping amplitudes to be $t_2=-0.03$~eV and
$t_1=0.8t_2$  which leads to a non-interacting bandwidth of $W
\approx 0.3$~eV, comparable to values from density functional
theory (DFT) calculations \cite{Merino00b}.

\begin{figure}
\begin{center}
\epsfig{file=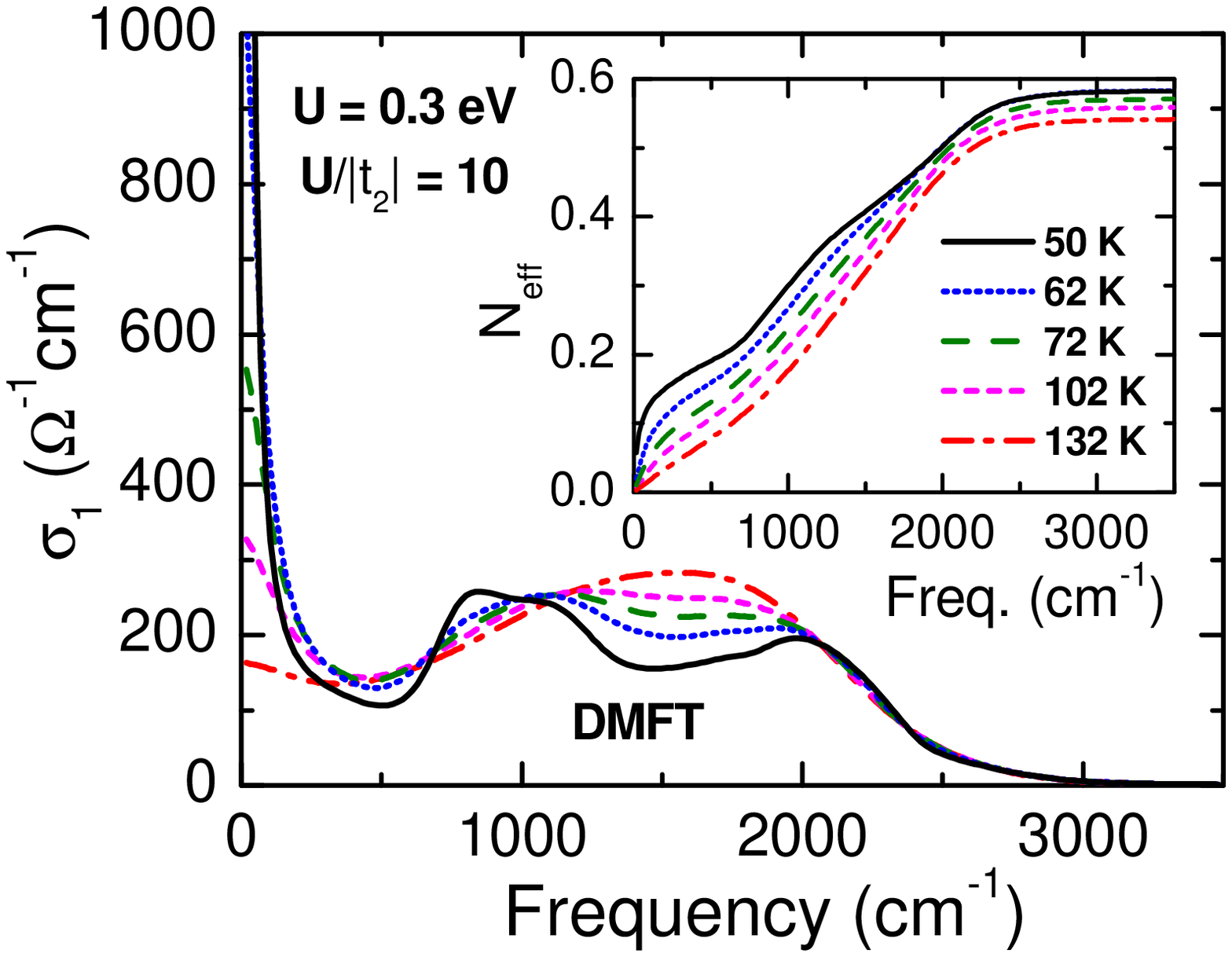,width=6.5cm,angle=0,clip=}
\end{center}
\caption{(Color online). Frequency dependence of the optical
conductivity at several temperatures. Calculations are based on a
DMFT treatment of the Hubbard model on the anisotropic triangular
lattice. For $U=0.3$ eV, a gradual suppression of the Drude peak
occurs with increasing temperature. Above the coherence
temperature, $T^* \approx 100$ K, the Drude peak disappears, due
to the destruction of Fermi liquid quasiparticles. The inset shows
the effective number of charge carriers $N_{\rm eff}(\omega)$ per
lattice site, calculated from the spectrum using Eqns.
(\ref{plasma}) and (\ref{neff}). } \label{fig1}
\end{figure}

Previous studies of model~(\ref{model}) based on DMFT calculations
\cite{Merino00a} and its cluster extensions \cite{Parcollet}
considered the Mott-Hubbard metal-insulator transition driven by
$U/W$. Anomalous transport properties observed in
$\kappa$-(BEDT-TTF)$_2X$ \cite{Limelette00}, such as the
non-monotonic temperature dependence of  the DC resistivity,
thermopower, and Hall coefficient agrees with DMFT, which predicts
the gradual destruction of quasiparticles with increasing
temperature, $T$. This leads to a crossover from Fermi-liquid
behavior at low $T$ to `bad' metallic behavior for $T>T^*$, with
$T^*<<W$ \cite{Merino00a}.
 Signatures of
the gradual destruction of quasiparticles as the temperature is
increased above $T^* \approx 100$~K can be seen in
Fig.~\ref{fig1}, which shows the frequency dependence of the real
part of the optical conductivity $\sigma_1(\omega)$ for several
temperatures. The results are obtained from DMFT using iterative
perturbation theory \cite{Kotliar96}. At the lowest temperature
three features are found: a Drude peak at $\omega=0$, a broad
absorption band at $\omega \approx 2000~{\rm cm}^{-1} \sim U$, and
a band at $U/2$. The broad band is a result of electronic
transitions between the Hubbard bands separated by
$U=10|t_2|=0.3$~eV, while the  band at $U/2$ results from
transitions between the quasiparticle peak and Hubbard bands.

Single crystals of \etbrcl\ are investigated by polarized
reflectivity spectroscopy in a broad frequency and temperature
range (for details see Ref.~\onlinecite{Faltermeier07}). The
optical response can be separated into contributions from
correlated charge carriers, intradimer charge-transfer
transitions, and vibronic    modes activated by these
charge-transfer transitions. While the spectra of all the
compounds are similar at high temperatures, a clear distinction
can be made below 50~K. A Drude peak is present for
 a Br content of $x=0.73, 0.85$, and 0.9 indicating metallic
behavior.
 In contrast, no Drude peak in seen
in the materials with $x=0$ and 0.4, consistent with a Mott
insulating ground state \cite{Faltermeier07}.
 In order to compare the measurements to DMFT calculations,
 we followed the procedure used in
Ref.~\onlinecite{Faltermeier07} to   subtract  the contributions
of intradimer electronic transitions (around 3400~\cm) and
vibrational features from the optical conductivity spectra, since
the associated physics is not incorporated in the model
(\ref{model}).

\begin{figure}
\begin{center}
\epsfig{file=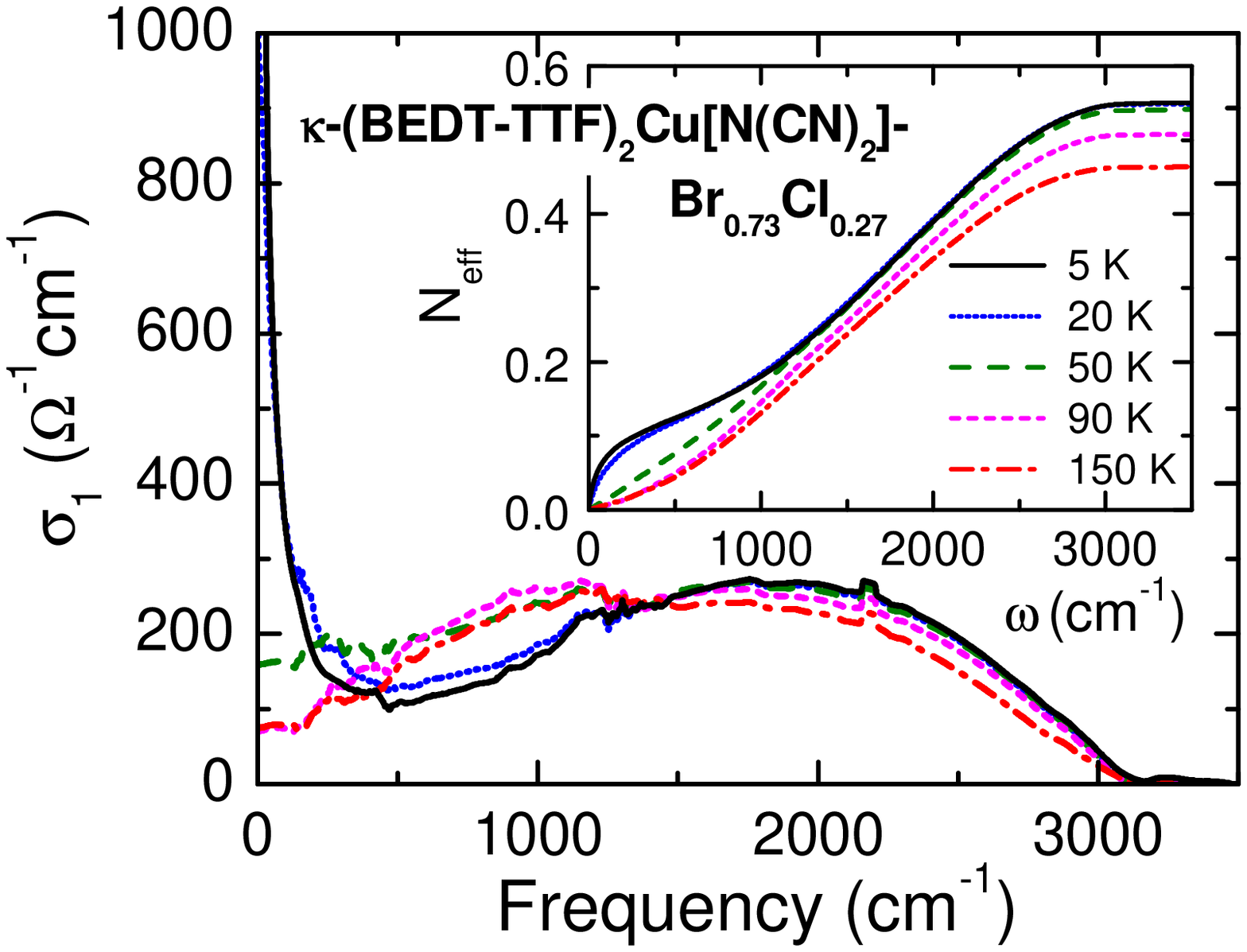,width=6.5 cm,angle=0,clip=}
\end{center}
\caption{(Color online). Frequency dependence of the optical
conductivity
 due to the
correlated charge carriers of
$\kappa$-(BE\-DT\--TTF)$_2$\-Cu\-[N\-(CN)$_{2}$]\-Br$_{0.73}$Cl$_{0.27}$.
Experimental data \cite{Faltermeier07} are shown for conductivity
in the layers
 ($E \parallel c$) at different temperatures.
This data compares favourably with theoretical calculations
presented in Fig.~\ref{fig1}. The inset shows the effective number
of charge carriers $N_{\rm eff}(\omega)$ per dimer. } \label{fig2}
\end{figure}

The optical conductivity, $\sigma_1(\omega)$, is shown in
Fig.~\ref{fig2} for the compound with 73\%\ Br. We confine
ourselves to the polarization $E~\parallel~c$; [the crystal
$c$-axis is parallel to the diagonal $k_y=k_x$ in the model
(\ref{tight})] similar results are obtained along the second
intralayer crystal axis. At $T=150$~K, one broad absorption
feature is observed centered around 2000 \cm. When the sample is
cooled down, the SW shifts towards lower frequencies and below
50~K a narrow Drude-like component develops. Interestingly, the
overall SW decreases for $T> 50$~K. The redistribution of spectral
weight with varying temperature, observed in Fig.~\ref{fig2} is
consistent with the calculations shown in Fig.~\ref{fig1}. The
agreement is particularly impressive as there are really only two
parameters in the calculation, the magnitude of $t_2$ (which sets
the vertical scale) and the ratio $U/t_2$ which determines the
relative weight in the Drude peak. However, we do note two
discrepancies. First, the band at $\omega \sim U/2$ does not
appear in the experimental data. Second, the temperature scale on
which the Drude weight collapses is about a factor of two larger
in the calculation than in the experiment. Such discrepancies are
not unusual in systems exhibiting Kondo physics, as occurs in DMFT
\cite{Kotliar96}.

To describe the redistribution of spectral weight with varying
temperature and correlation strength, it is useful to consider
restricted f-sum rule \cite{Marel05,Maldague} based on the
integral:
\begin{equation}
I_\sigma(\Lambda)\equiv \int_0^\Lambda \sigma_1(\omega)\, {\rm
d}\omega \equiv \frac{\pi n e^2}{2 m^*_{\rm sum}} \label{fsum}
\end{equation}
where $\Lambda$ is a high-frequency cut-off which excludes
interband transitions (we take $\Lambda=3500$ \cm), $e$ is the
electronic charge, $n$ is the total electron density (excluding
filled bands) and the last identity defines a particular effective
mass $m^*_{\rm sum}$. Expression (\ref{fsum}) can be compared to
the full f-sum rule which  is independent of temperature and the
strength of the correlations \cite{DresselGruner02,Marel05}. When
the conductivity is integrated up to infinity, it can be related
to the plasma frequency of an electron gas with the total electron
density.

In order to make more detailed statements we have to consider the
effective mass and effective carrier concentration separately. For
a single two-dimen\-sion\-al metallic band of {\it
non-interacting} electrons all the spectral weight is in the Drude
peak and the sum rule is
\begin{equation}
I_\sigma(\Lambda)=
 {\pi  e^2 \over b}
 \int_{occ} {{\rm d}^2 {\bf k} \over (2 \pi \hbar)^2 }
{ \partial^2 \epsilon({\bf k}) \over \partial k_c^2} \equiv {\pi n
e^2  \over 2 m_{b,{\rm opt}} } \quad , \label{plasma}
\end{equation}
where 
$b$ is the interlayer spacing, and integral \ref{fsum} is over all
the occupied states (i.e.\ the Fermi sea).
 The last equality defines an optical band mass,
$m_{b,{\rm opt}}$ \cite{mass}. For the tight-binding model
(\ref{tight}) at half-filling, evaluating Eq.~(\ref{plasma}), with
$t_2=-0.03$~eV and $t_1=0.8t_2$ yields
 an optical band mass, $m_{b,{\rm opt}}=2.3m_e$.
This agrees very well with $m_{b,{\rm opt}}=2.5m_e$ obtained from
our experimental data \cite{Faltermeier07,Dumm08}.

For a Hubbard model with {\it only nearest-neighbor} hopping on an
isotropic lattice this integral is proportional to the expectation
value of the kinetic-energy operator
\cite{Marel05,Maldague,Rozenberg}. This sum rule even holds in the
absence of a Drude response and in the Mott insulating phase. The
total spectral weight -- as calculated in Eq.~(\ref{fsum}) -- can
then vary with temperature and the strength of the electronic
correlations \cite{Rozenberg}. However, we should point out that
although this connection with the kinetic energy will hold
qualitatively for our Hubbard model, it will not hold
quantitatively because the system of interest is spatially
anisotropic. In the Hubbard model (\ref{model}) there are
next-nearest neighbor terms ($t_1$) which are comparable to the
nearest neighbor terms and frustrate the kinetic energy.

\begin{figure}
\begin{center}
\epsfig{file=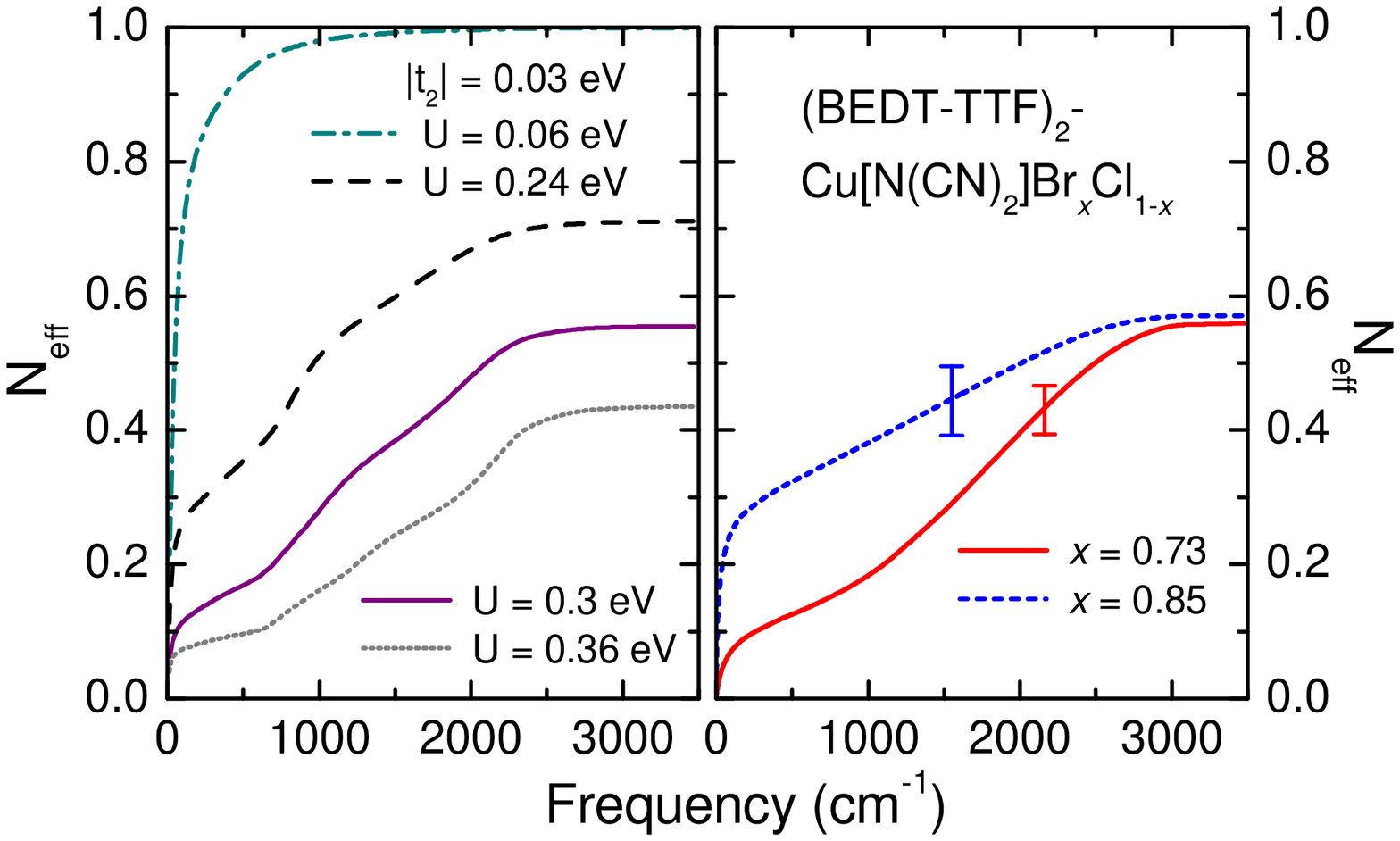,width=9cm,angle=0,clip=}
\end{center}
\caption{(Color online). Effective charge-carrier number per
(BEDT-TTF)$_2$ dimer approaching the Mott transition. In the left
panel the results of a DMFT calculation of $N_{\rm eff}(\omega)$
(at a temperature of 50K) show how the spectral weight is
redistributed as one approaches the transition to the Mott
insulating phase which occurs at $U=15|t_2|=0.45$ eV.
 The SW redistribution and saturation of
 $N_{\rm
eff}(\omega)$ is in agreement with the experimental data for
$\kappa$-(BEDT-TTF)$_2$Cu[N(CN)$_2$]Br$_x$Cl$_{1-x}$ for $x =0.73,
0.85 $ (at low temperature)
 shown in the right panel.
Typical error bars for the experimental data are shown. }
\label{fig3}
\end{figure}
\begin{figure}
\begin{center}
\epsfig{file=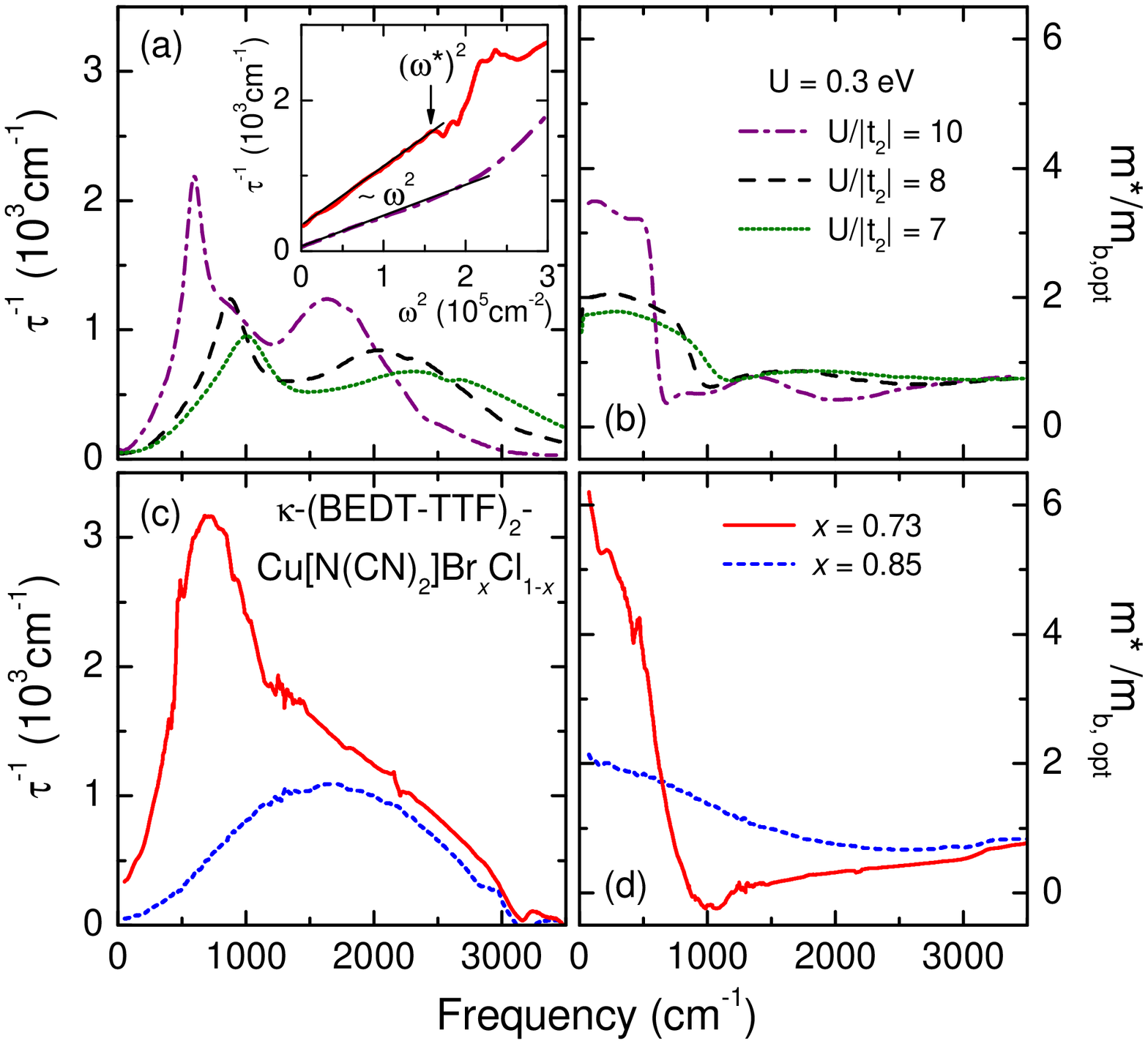,width=9cm,angle=0,clip=}
\end{center}
\caption{(Color online). Frequency dependence of the scattering
rate and effective mass  extracted from an extended
 Drude model analysis of the optical conductivity.
(a) and (b) shows the results of DMFT calculations
 for different $U/W$ and $T=50$K.
(c) and (d) show the experimental results for the same quantities
at $T=5$K. The inset of (a) shows how below a frequency
 $\omega^* \approx 400$~\cm\ the scattering
rate found for both experiment and theory has the quadratic
frequency dependence expected for Fermi liquid quasiparticles.
Clearly as the Mott insulating phase is approached the effective
mass and the scattering rate increase significantly.} \label{fig4}
\end{figure}

For the further analysis of the effects of electronic correlations
on the charge-carrier kinetic energy, we rewrite Eq.~(\ref{fsum})
and define a frequency-dependent carrier density:
\begin{equation}
N_{\rm eff}(\omega)={2 m_{b,{\rm opt}}  \over \pi e^2}
 \int_0^\omega
\sigma_1(\omega')\,{\rm d} \omega' \quad .
 \label{neff}
\end{equation}
where at infinite frequencies $N_{\rm eff}$ is equal to the number
of charge carriers on a dimer (f-sum rule), while deviation from
this value indicate the change in kinetic energy; i.e.\ the
influence of electronic correlations.
For a weakly correlated metal, $N_{\rm eff}(\omega)$ increases
rapidly from zero to unity once $\omega$ becomes larger than the
intraband scattering rate, because almost all the spectral weight
is in the Drude peak (see the $U=0.06$ eV curve in
Fig.~\ref{fig3}). As $U/W$ is increased, both in theory and
experiment (Figs.~\ref{fig3} and \ref{fig4}), SW is transferred to
higher frequencies and $N_{\rm eff}(\omega)$ saturates above
frequencies of about $\omega \sim U$. The absolute value of
$N_{\rm eff}$ is substantially suppressed. Qualitatively, the
reduction in the total spectral weight with increasing
correlations is due to the diminishing kinetic energy of the
charge carriers.

In the inset of Fig.~\ref{fig1}, $N_{\rm eff}(\omega)$ evaluated
using DMFT with $U=0.3$ eV is plotted for different temperatures.
At large frequencies, the integrated spectrum saturates to about
0.6. The calculations are consistent with experimental results for
$\kappa$-(BE\-DT\--TTF)$_2$\-Cu\-[N\-(CN)$_{2}$]\-Br$_{0.73}$Cl$_{0.27}$,
shown in the inset of Fig.~\ref{fig2} and the right panel of
Fig.~\ref{fig3}. With increasing temperature, $N_{\rm
eff}(\omega)$ is suppressed at low frequencies and spectral weight
is transferred to high $\omega$, whereas the total spectral weight
of the correlated carriers is not conserved because of the gradual
destruction of quasiparticles with increasing temperature.

For a more sophisticated analysis of the mass enhancement of the
correlated carriers, the frequency dependent mass renormalization
can be quantified with the generalized Drude model
\cite{Puchkov,DresselGruner02}:
\begin{equation}
\sigma(\omega)= \sigma_1(\omega)+i\sigma_2(\omega)=
{(2/\pi)I_\sigma(\Lambda) \over 1/\tau(\omega) - i \omega
m^*(\omega)/m_{b,{\rm opt}}} \quad , \label{drude}
\end{equation}
where $m^*(\omega)$ and $1/\tau(\omega)$ are a frequency dependent
optical effective mass
 and scattering rate, respectively.
At high frequencies the effective mass becomes equal to the
optical band mass, since at high frequencies, $\sigma(\omega)$ is
dominated by the imaginary part.
The frequency dependences of this scattering-rate and
effective-mass enhancement
 are plotted in Fig.~\ref{fig4}.
  A sharp drop in $m^*(\omega)$ and a peak in
$1/\tau(\omega)$ both occur in DMFT at a frequency, $\omega_c$ as
shown in Figs. 4(a) and 4(b). Below a  frequency scale, $\omega^*
$
  Fermi liquid behavior, {\it i.e.},
$1/\tau(\omega) \propto \omega^2$, occurs. For $U=0.3$ eV, the
DMFT calculations give: $\omega^* \approx 400$ cm$^{-1}$ (inset of
Fig. \ref{fig4}a) in good agreement with the experimental value.
Both $\omega_c$ and $\omega^*$ are shifted to lower frequencies
with increasing $U/W$, consistent with the shift observed as $x$
decreases from 0.85 to 0.73. The collapse of the effective mass
enhancement at high frequencies is reminiscent of what occurs in
heavy fermion compounds\cite{degiorgi} and the `kink' seen in
ARPES experiments on cuprate superconductors and has recently been
argued to be a ubiquitous feature of strongly correlated electron
systems \cite{Vollhardt,Prelovsek}.

The large increase in the effective optical mass as the system
comes closer to the Mott insulating phase (due to increasing Cl
content) is consistent with the large mass renormalization
expected from Brinkman-Rice and DMFT pictures\cite{Kotliar96}.
This is distinctly different from what occurs in doped Mott
insulators, such as the cuprates, for which the effective mass
deduced from an extended Drude model analysis is  weakly dependent
on the doping as the Mott insulating phase is approached
\cite{Uchida,Haule}. For $U/t_2=2,8,10,12$, the enhancements in
the effective mass of the Fermi liquid quasiparticles calculated
from DMFT are $m^*/m_{b,{\rm opt}}=1.1, 2.1, 3.1,$ and 4.1,
respectively.

In conclusion, quasiparticles at the verge of localization in
$\kappa$-(BEDT-TTF)$_2$Cu[N(CN)$_2$]Br$_x$Cl$_{1-x}$ display a
sizeable redistribution of spectral weight as the temperature
and $x$ vary. The strong local Coulomb correlations developing
close to the Mott transition at $x =0.7$ lead to spectral-weight
transfer from low to high frequencies. The effective masses
extracted from optical data are strongly enhanced approaching the
Mott transition, in contrast to doped Mott insulators. These
effects lead to the destruction of coherent excitations above a
frequency scale, $\omega^*$.

We thank P. Batail, C. Meziere, D. Faltermeier, and B.J. Powell
for their contributions to this work. The project was supported by
the Ram\'on y Cajal program from MCyT in Spain, the MEC under
contract CTQ2005-09385-C03-03, the grant `Leading Scientific
Schools' NSh-5596.2006.2, the Deutsche Forschungsgemeinschaft, and
the Australian Research Council.

\end{document}